\documentclass{aastex}
\usepackage{longtable}
\usepackage{natbib}
\usepackage{amssymb,amsmath}

\shorttitle{Photometric and Spectroscopic Observations of  GRB 140629A}
\shortauthors{Xin et al. }

\begin{document}

\title{Photometric and Spectroscopic Observations of  GRB 140629A }

\author{Li-Ping Xin \altaffilmark{1}, Shu-Qing Zhong \altaffilmark{2}, En-Wei Liang \altaffilmark{2}, Jing Wang \altaffilmark{1,4}, Hao Liu \altaffilmark{1}, Tian-Meng Zhang \altaffilmark{3}, Xiao-Li Huang \altaffilmark{2}, Hua-Li Li \altaffilmark{1}, Yu-Lei Qiu \altaffilmark{1},  Xu-Hui Han\altaffilmark{1}, Jian-Yan Wei \altaffilmark{1,4} }
\altaffiltext{1}{CAS Key Laboratory of Space Astronomy and Technology, National Astronomical Observatories, Chinese Academy of Sciences, Beijing 100012, China. xlp@nao.cas.cn}
\altaffiltext{2}{Guangxi Key Laboratory for Relativistic Astrophysics, School of Physical Science and Technology, Guangxi University, Nanning 530004, China; lew@gxu.edu.cn}
\altaffiltext{3}{Key Laboratory of Optical Astronomy, National Astronomical Observatories, Chinese Academy of Sciences, Beijing 100012, P.R. China}
\altaffiltext{4}{School of  Astronomy and Space Science, University of Chinese Academy of Sciences, 19A Yuquan Rd, Shijingshan District, Beijing, P.R.China 100049}
\begin{abstract}
We present our optical photometric and spectroscopical observations of GRB 140629A. A redshift of $z=2.275\pm0.043$ is measured  through the metal absorption lines in our spectroscopic data. Using our photometric data and multiple observational data observed with other telescopes, we show that its optical light curve is well interpreted with the standard forward shock models in the thin shell case. Its optical-X-ray afterglow spectrum is jointly fitted with a single power-law function, yielding a photon index of $-1.90\pm 0.05$. The optical extinction and neutral hydrogen absorption of the GRB host galaxy are negligible. The fit to the light curve with the standard models shows that the ambient density is $60\pm 9$ cm$^{-3}$ and the GRB radiating efficiency is as low as $\sim 0.24\%$, likely indicating  a baryonic-dominated ejecta of this GRB. This burst agrees well  with the $L_{\rm p, iso}-E_p^{'}-\Gamma_0$ relation, but confidently violates those empirical relations involving geometric corrections (or jet break time). This gives rise to an issue of  possible selection effect on these relations since the jet opening angle of this GRB is extremely narrow (0.04 rad).
\end{abstract}

\keywords{Gamma-ray busts --- stars: individual (GRB 140629A)---techniques: photometric--- techniques: spectroscopic}

\section{Introduction}
Gamma-ray bursts (GRBs) and their afterglows in soft energy bands are the most luminous events in the deep universe (M\'{e}sz\'{a}ros 2006; Kumar \& Zhang 2015). Typically, their short gamma-ray flashes may release an amount of isotropic energy in the gamma-ray band ($E_{\rm \gamma, iso}$) of $10^{50}-10^{54}$ ergs within tens of seconds. Their optical emissions may be so bright that some of them can be even occasionally seen by naked eye as GRB 080319B (Racusin et al. 2008). As an expectation of collimated jet models (Rhoads 1999; Harrison et al. 1999; Dai et al. 2007 ), jet breaks have been detected in the late multi-wavelength afterglow light curves of some bursts (e.g., Nicuesa et al. 2011), which make their true energy release  being smaller than the isotropic one by 2 to 3 orders of magnitude (e.g., Frail 2001; Bloom et al. 2003).

The discovery of the multi-wavelength emission of the afterglows has revolutionized our understanding on the GRB phenomenon (e.g., Piran 1999; Zhang \& M{\'e}sz{\'a}ros 2004). Due to the rapid response and the precise localization capabilities of the X-ray telescope (XRT) on board the {\em Swift} mission, X-ray afterglows are detected for more than 96\% of the GRBs which trigger the {\em Swift} Burst Alert Telescope (BAT; Burrows et al. 2007). Most of the well-sampled XRT light curves usually start with bright flares and/or a steep decay segment with a slope\footnote{ The flux convention $F\propto t^{\alpha} \nu^{\beta}$ is adopted, where $\alpha$ and $\beta$ are the temporal and spectral indices, respectively.} of $\alpha<-3$ (Zhang et al. 2006; Noseck et al. 2006; O'Brien et al. 2006). The joint spectral analysis of these X-ray flares with simultaneous gamma-ray pulses indicate that they are the low energy extension of the prompt gamma-ray emission (Peng et al. 2014). The initial steep decay phase is rather explained as being the tail emission of the last gamma-ray emission pulse due to the so-called curvature effect (e.g., Liang et al. 2006; Zhang et al. 2007; Mu et al. 2016). Following the initial steep decay segment, XRT light curves usually have a shallow decay segment with a slope of $\alpha\sim -0.5$ or even shallower before transferring  to the so-called standard decay segment with a slope of $\alpha\sim -1$. These features well agree with the predictions of the external shock models with extra energy injection (Dai \& Lu 1998; Zhang \& M{\'e}sz{\'a}ros  2002; Liang et al. 2007). A jet-like decay segment is only observed in a few XRT light curves (Liang et al. 2008), and a small fraction of XRT light curves are featureless showing a single power-law flux decay from very early to late epochs (Liang et al. 2009).


Large sample analysis for GRB optical light curves  (e.g., Oates et al. 2009; Kann et al. 2010; Panaitescu \& Vestrand 2011; Li et al. 2012) shows that a significant fraction of optical light curves were found to start with a smooth bump  and then decay as a power-law decay. This feature might be due to the deceleration of the GRB fireball by the surrounding medium as predicted by forward shock models in the thin shell case (Sari \& Piran 1999; Kobayashi \& Zhang 2007).  About one-third of their optical light curves start with a shallow decay segment, as usually seen in XRT light curves (Li et al., 2012).
Detection rate of optical flares is much lower than that of X-ray flares (e.g. Li et al. 2012; Swenson et al. 2013). It was proposed that early optical light curves may be good probes for investigating the properties of fireballs and the ambient density (e.g., Liang et al. 2010, 2013; Yi et al. 2013; Xin et al., 2016a). Although their chromatic breaks observed in both X-ray and optical afterglow light curves give rise to an issue for explaining their physical origins (Panaitescu et al. 2006; Fan et al. 2006; Liang et al. 2007), the X-ray and optical data can be accommodated within the external shock models by considering various effects (e.g., Cucchiara et al. 2011; Wang et al. 2015).

In this paper, we report our optical photometric and spectroscopical observations of GRB 140629A with the TNT telescope and the 2.16 m telescope at Xinglong Observatory. We show that these observations are consistent with  the prediction of the external shock fireball model in the thin shell case. Combining our data with other observations carried out by Swift, Konus-Wind, and other ground based telescopes, we derive the jet properties of this GRB and examine whether it satisfies empirical relations derived from observations of both the prompt gamma-rays and the afterglows. Our observations and data reduction are presented in \S 2. The analysis of the spectroscopical data and the redshift measurement of GRB 140629A are reported in \S 3. The analysis of the optical and X-ray afterglow photometric  data and our modelling of the afterglow light curves are reported in \S 4. A discussion of our results and our conclusions are presented in \S 5 and \S 6. A standard cosmology model with $H_0=70$ km/s/Mpc, $\Omega_M=0.27$, and $\Omega_{\Lambda} = 0.73$ is adopted. The notation $Q_n=Q/10^{n}$ is in cgs units.

\section{Observations and data reduction}
\subsection{Prompt Gamma-ray emission and X-Ray afterglow Observations}
GRB 140629A triggered the Swift Burst Alert Telescope (BAT) at 14:17:30 UT on June 29, 2014 ($T_0$; Lien et al. 2014). It was also detected by Konus-Wind in the waiting mode (Golenetskii et al. 2014). The BAT light curve starts with weak and overlapped emission peaks prior to the BAT trigger of about 8 seconds and ends at about $90$ seconds post trigger time with some fluctuations (Cumming et al. 2014), as observed in some long BAT GRBs (Hu et al. 2014). The burst duration ($T_{90}$) is $42.0\pm 14.3$ s measured in the BAT 15 -150 KeV energy band  (Cumming et al. 2014). Its duration measured by Konus-Wind in the energy range from 20 to 10000 KeV is less than 26 sec (Golenetskii et al. 2014), being much shorter than that in the BAT band. This confirms that the duration of the GRB prompt emission depends on the instrument energy band (e.g., Qin et al. 2013). The time-integrated prompt gamma-ray spectrum observed with Konus-Wind in the $20-10^4$ keV band can be best fitted with a cutoff power-law function, yielding a photon index of $\Gamma_\gamma= -1.42\pm 0.54$ and a peak energy of the $\nu f_\nu$ spectrum of $E_p = 86\pm 17$ keV. The associated gamma-ray fluence and the peak flux are $S_\gamma=(3.4\pm0.5)\times 10^{-6}$ erg cm$^{-2}$ and $F_\gamma=(4.7 \pm 0.7)\times 10^{-7}$ erg cm$^{-2}$ s$^{-1}$, respectively (Golenetskii et al. 2014). The X-ray afterglow was detected by XRT at a time $t>T_0+93$ seconds, roughly at the end of the prompt emission. We obtain the BAT and the XRT light curves from the XRT light curve and spectral repository (Evans et al. 2007, 2009)\footnote{ http://www.swift.ac.uk/}. As shown in the Figure \ref{obs_lc}, the prompt gamma-rays show two episodes. The first episode lasts from $T_0-8$ seconds to $T_0+10$ seconds. The second episode is from $T_0+10$ seconds to $T_0+90$ seconds. 

\subsection{Photometric Observations of the Optical Afterglows}
The bright optical counterpart of GRB 140629A was detected  by several ground-based telescopes, such as the three MASTER system telescopes located in Blagoveshchensk, Tunka and Kislovodsk (Yurkov et al. 2014; Gorbovskoy et al. 2014), the Russian-Turkish 1.5-m telescope (Bikmaev et al. 2014), the 1.05-m Schmidt telescope at Kiso Observatory in Japan (Maehara et al. 2014), the Murikabushi 1-m telescope of Ishigakijima Astronomical Observatory (Kuroda et al. 2014), the Nordic Optical Telescope (NOT), and the Palomar 60-inch (P60) robotic telescope (Perley et al. 2014).

Our optical follow-up observation campaign of GRB 140629A was carried out using the TNT (0.8-m Tsinghua University - National Astronomical Observatory of China Telescope) at Xinglong Observatory, beginning at $T_0+581$ seconds and ending at about 2.15 hours after the {\em Swift}/BAT trigger time, respectively. Several $B$, $V$, $R$ and $I$-band images were obtained. The data reduction was carried out following the standard routine in IRAF\footnote{IRAF is distributed by NOAO, which is operated by AURA, Inc., under cooperative agreement with NSF.} package, including bias and flat-field corrections. Dark correction was not performed since its impact on the source extraction and photometry is  negligible once  the CCD was cooled down to $-110\,^{\circ}\mathrm{C}$. A point spread function (PSF) photometry method was applied via the DAOPHOT task in the IRAF package. During the reduction, $B$-band frames were stacked in order to increase the signal-to-noise (S/N) ratio. An absolute photometric calibration was performed using the Sloan Digital Sky Survey (SDSS, Adelman-McCarthy et al. 2008), with flux/mag conversion of the SDSS system into the Johnson-Cousins system\footnote{http://www.sdss.org/dr6/algorithms/sdssUBVRITransform.html \#Lupton2005}.
All the data we obtained by TNT are presented in Table \ref{Tab1}.
For more details of the follow-up system of TNT and the data reduction please refer to Zheng et al., (2008) and Xin et al. (2011).

A well-sampled optical light curve is obtained from our observations from $T_0+580$ seconds to $T_0+2.15$ hours. In order to get an optical light curve in broader temporal coverage, we collect the early and late optical observations of the other telescopes from GCN Circulars (Malesani et al. 2014; Bikmaev et al. 2014; Masi. 2014; Sonbas et al. 2014; Moskvitin et al. 2014b,2014c; Perley \& Cenko 2014; Gorbovskoy et al. 2014). Note that the early optical data observed with the MASTER system telescopes were also re-calibrated to  USNO B1.0 R2 mag as done for our observations.  During $t\sim 600-800$ sec after the burst trigger, simultaneous observations with the MASTER system and our TNT telescope are available.
We find that the corrected magnitudes derived from the data observed with the MASTER system are systematically brighter than TNT data by $\Delta R=0.51$ mag. The discrepancy might be caused by the flux calibration between the two telescopes. We therefore re-normalized the MASTER data to the TNT data by adding $\Delta R=0.51$ mag  based on the simultaneously observed data during $t\sim600-800$ sec after the burst.  We finally obtain an optical light curve covering a long duration from $T_0+38$ seconds to $T_0+1.4\times 10^5$ seconds, as shown in Figure \ref{obs_lc}.

\subsection{Spectroscopic observations of the optical afterglows}
We carried out spectroscopic observations with the National Astronomical
Observatories, Chinese Academy of Sciences (NAOC) 2.16 m telescope (Fan et al., 2016) in Xinglong Observatory  on
2014 June 29 at 15:10:48 (UT), about one hour after the burst trigger.
The optical spectrum was obtained with an Optomechanics Research Inc. spectrograph. The spectrograph is
equipped with a back-illuminated SPEC 1340$\times$400 CCD.
The grating is 300$\mathrm{g\ mm^{-1}}$, and the slit oriented
in the south-north direction corresponds to a width of
2.\symbol{125}0. This setup finally results in a spectral resolution of
$\sim$9\AA, as measured from the sky emission lines and comparison arcs.
The spectrum was blazed at a wavelength of 6000\AA, and was obtained with an exposure time of 2400s.
The two-dimensional spectrum was reduced by the standard
procedures through the IRAF package, including bias subtraction, flat-field correction, and cosmic ray removal
before the extraction of the one-dimensional spectrum.
The extracted one-dimensional spectrum was then calibrated in wavelength by the helium-neon-argon comparison arc taken
immediately after the exposure. The subsequent resulting wavelength accuracy is better than 1\AA. The calibration in flux was carried out by
the Kitt Peak National Observatory (KPNO) standard stars BD+332642 (Massey et al. 1988).
The two telluric features at around $\lambda$6800 and $\lambda$7600 due to $\mathrm{O_2}$ molecules were removed from
the observed spectrum by the standard calibration stars.

\section{ Redshift measurement and optical spectrum features}
\begin{figure}
\centering
\includegraphics[angle=0,scale=0.8]{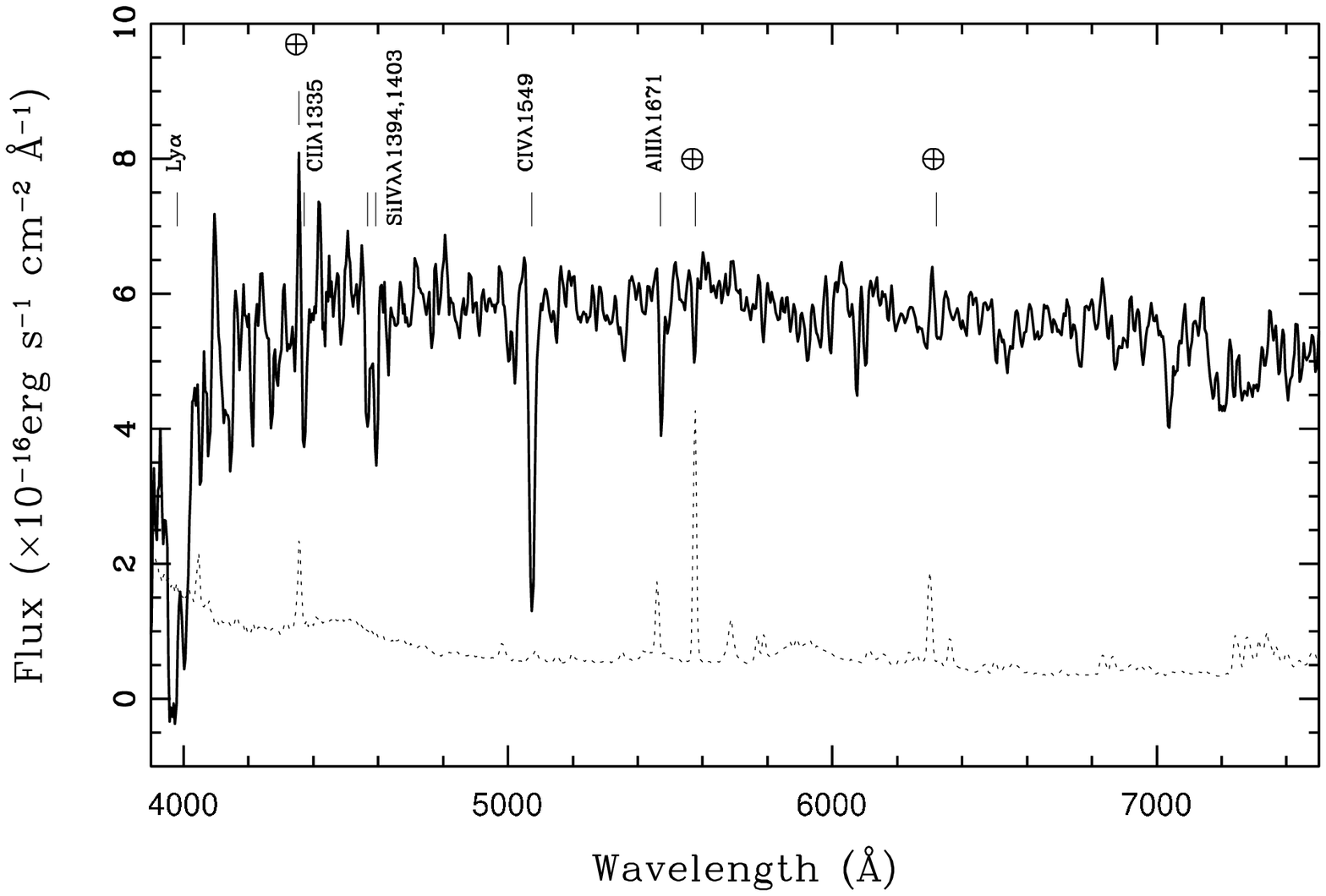}
\includegraphics[angle=0,scale=0.8]{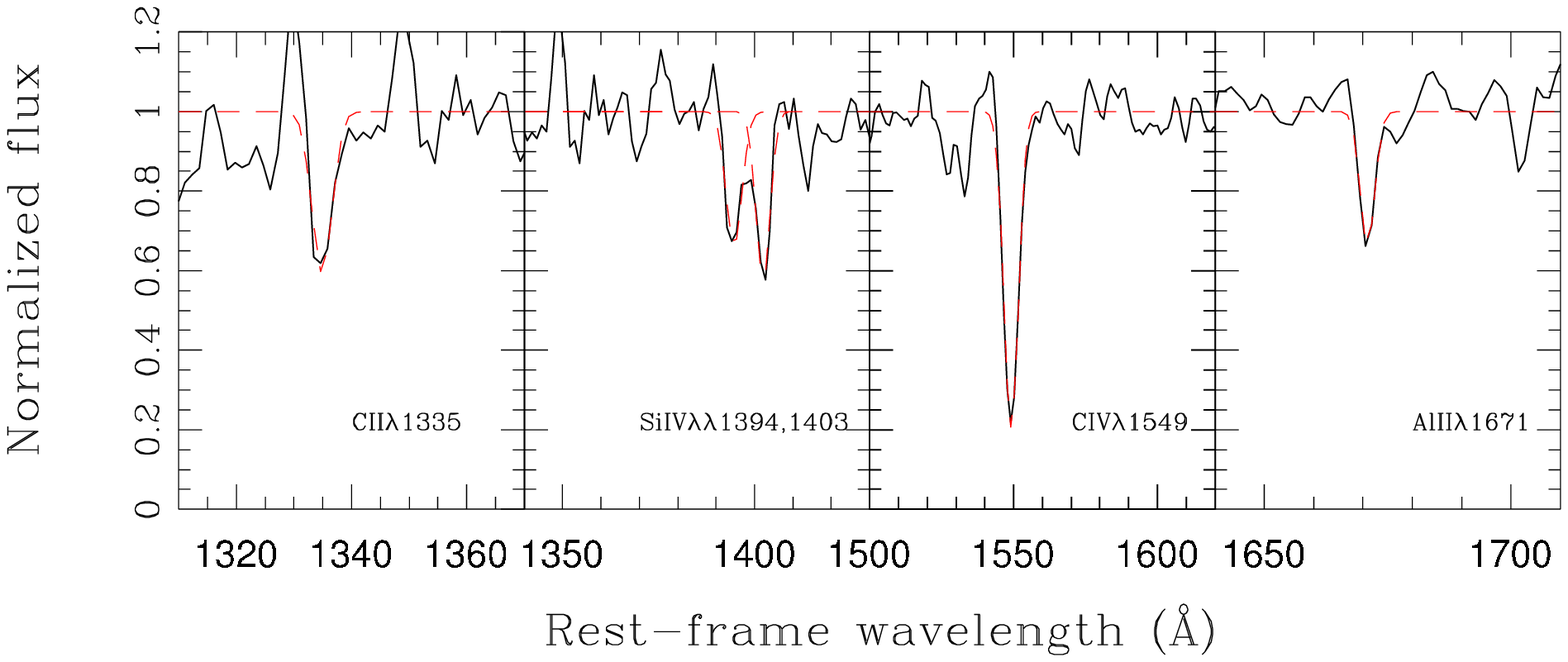}
\caption{{\em Upper---}The optical spectrum of GRB 140629A (the solid curve) obtained with Xinglong 2.16-m
telescope at about 1 hour after the burst trigger. The spectrum is smoothed by a box size of 3\AA.
The dashed curve illustrates the night sky emission spectrum. The identified
features both from the GRB afterglow and from the poor subtraction of the night sky emission are marked.  {\em Bottom---}Line modelings with a Gaussian function for the five identified absorption features in the rest-frame. In each panel,
the normalized observed spectrum and the best fit model are shown by the black solid line and by the red dashed line, respectively.}
\label{Spectroscopy}
\end{figure}

In order to enhance the signal-to-noise ratio, the spectrum was smoothed by a box size of 3\AA. The reduced spectrum in the observer frame is shown in Figure \ref{Spectroscopy}. By excluding the artificial features due to the poor subtraction of the night sky emission, 
a series of hydrogen and metal absorption ( Ly$\alpha$, CII$\lambda$1335, SiIV$\lambda\lambda$1394,1403,
CIV$\lambda1549$ and AlII$\lambda$1671 ) are identified from the optical spectrum.
The redshift of GRB 140629A is determined through the metal absorptions because of the damped Ly$\alpha$ absorption and the poor signal-to-noise ratio at the blue end. We finally obtain a redshift of $z=2.275\pm0.043$,
in which the wavelength of each line center in the observer frame is estimated by a line
profile modeling method using a Gaussian function. Our result is consistent with the reported by other
groups (Moskvitin et al. 2014a; D'Avanzo et al. 2014).

Based on our estimated redshift, we derive the optical spectrum in the rest-frame.
The correction for the Galactic extinction has been applied using a colour excess $E(B-V)$ taken from the NASA/IAPC Extragalactic
Database (NED), assuming  $R_V$=3.1 (Cardelli et al. 1989).
We model each absorption feature in the rest-frame by a Gaussian profile through the IRAF/SPECFIT package task (Kriss 1994),
except for the damped Ly$\alpha$ absorption. Our results are illustrated in Figure \ref{Spectroscopy}.
The measured equivalent widths (EWs) in the rest-frame are reported in Table 2. All the uncertainties given Table 2
only include the statistical errors resulting from the spectral fitting.
One can observe that
the strongest metal absorption occurs in CIV$\lambda1549$, which is consistent with previous statistical studies based on low
resolution afterglow spectroscopy (e.g, de Ugrate Postigo et al. 2012). The ratio between CIV$\lambda1549$ and
CII$\lambda$1335 is a good indicator of ionization by the GRB's intense radiation. The inferred ratio
$\mathrm{EW(CIV)/EW(CII)}$ is $2.79\pm0.49$, which is higher than the reported average value by a factor of 2 ( see Table 8 in
de Ugrate Postigo et al. 2012). The variation of the fine-structure of the ions has been
already observed in several GRBs (e.g., Vreeswijk et al. 2007; D'Elia et al. 2009).
Dessauges-Zavadsky et al. (2006) reported a
significant decrease of FeII$\lambda$2396 transition by a factor of 5 in the afterglow spectrum of GRB\,020813.
The high ionization revealed in the early afterglow spectrum of GRB 140629A one hours after the burst trigger could be
due to a temporal evolution of the ionization as long as the GRB afterglow radiation.


\section{Optical and X-ray afterglow data analysis}
\subsection{Temporal analysis}
\begin{figure}
\centering
\includegraphics[angle=0,scale=0.4]{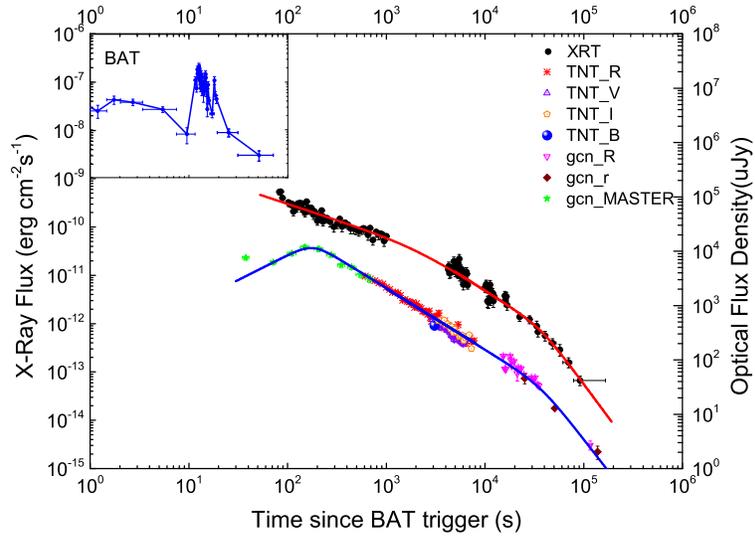}
\caption{The multi-wavelength light curves of GRB 140629A and our empirical fits to the R band and X-ray light curves with smooth broken power-laws.}
\label{obs_lc}
\end{figure}

Figure \ref{obs_lc} shows the multi-wavelength light curves of GRB 140629A.
The first optical data before 650 sec after the burst trigger time  could be attributed to the prompt emission in the optical band or the reverse shock emission,
 similar to that observed in GRB 140512A (Huang et al. 2016). We exclude this data in our following analysis.

We fit the optical and X-ray afterglow light curves with a multiple broken power-law model. Each broken power-law function is described as (Beuermann et al., 1999)
\begin{equation} F=F_0\left [
\left (   \frac{t}{t_b}\right)^{\omega\alpha_1}+\left (
\frac{t}{t_b}\right)^{\omega\alpha_2}\right]^{1/\omega},
\end{equation}
where $t_b$ is the break time, $\alpha_1$ and $\alpha_2$ are decay indices before and after the break, respectively, and $\omega$ describes the sharpness of the break, which is fixed as 3 in our analysis. Our empirical fits are illustrated in Figure \ref{obs_lc}, and summarized in Table \ref{Tab3}.

Note that by analyzing the UVOT data of 27 {\em Swift} GRBs, Oates et al., (2009) report that three GRBs which show clear bumps in their UVOT light curves. The rising slopes of these three GRBs are in the range from 0.26$\pm$0.13 to 0.73$\pm$0.14 before 500 sec after the bursts. The UVOT light curves decays with a slope ranging from $-0.5\pm0.05$ to $-1.67\pm0.15$ after 500 sec post the BAT trigger. They proposed that the rising in the optical light curves may be attributed to either the start of the forward shock , or to an off-axis viewing angle where the observer sees an increasing amount of emission as the Lorentz factor of the jet decreases. By analyzing a sample of 17 GRBs with early bumps in their early optical light curves, Liang et al. (2010) showed that the peak time of the early bump is in the range of $10^2-10^3 $ sec with a median value of $\sim$380 sec, and their rising slopes $r$ is in the range of $1 - 2$, expect for three exceptional GRBs, GRB 080330A with $r\sim0.34$, GRB 060607A with $r\sim4.15$ and GRB 050820A with $r\sim4.45$. Their decay slopes are distributed in the range of 0.44-1.77, with an average of 1.16$\pm$0.34. Liang et al. (2010) suggested that these bumps could be interpreted as the onset of the forward shock emission and the peak time is the deceleration time of the fireball. For GRB 140629A, we have $\alpha_{\rm O,1}=0.92\pm 0.24$, $t_{\rm O, p}=179\pm 16$s, and $\alpha_{\rm O,2}=-1.12\pm 0.02$,  indicating that the early smooth optical peak could be also attributed to the afterglow onset when the GRB fireball is decelerated by the ambient medium (e.g., Sari et al. 1999; Liang et al. 2010). The optical light curve transits to a steeper segment with $\alpha_{\rm O, 3}=-2.35\pm 0.24$ at $t_{\rm O, j}\sim 37$ ks.

The X-ray afterglow light curve starts with a shallow decay segment with a slope of $\alpha_{\rm X,1}=-0.67\pm 0.02$ up to $t_{\rm X, b}\sim 2$ ks, which very smoothly transits to a decay slope of $\alpha_{\rm X,2}=-1.31\pm 0.08$ until a break at $t_{\rm X,j}=37.2\pm 9.1$ ks. The decay slope post $t_{\rm X, j}$ is $\alpha_{\rm X,3}=-2.76\pm 0.40$. The decaying behavior of the X-ray light curve post $t>200$ seconds is consistent with the optical light curve.

Achromatic breaks in the optical and X-ray bands are usually suspected to be produced by the jet effect (Rhoads 1999) or the end of energy injection (e.g., Dai \& Lu 1998; Liang et al. 2007). A jet break is featured as transition from a normal-decay segment with a slope of $\sim -1$ to a steep-decay slope with  slope of $\sim -2$. An energy injection break is usually illustrated as transition from a shallow decay segment with a slope of $\sim -0.5$ to a normal-decay segment with a slope of $\sim -1$. The achromatic break of the optical and X-ray afterglow lightcurves of GRB 140629A at $\sim37$ ks is consistent with the jet break scenario. The decay slope post the break time depends on the index of the electron energy distribution $p$. We have $\alpha_{\rm O,3}=-2.35\pm0.24$ and $\alpha_{\rm X,3}=-2.76\pm0.40$, likely suggesting a steep electron spectrum. The large change in slopes around the break time, i.e., $|\Delta\alpha_O|\sim1.23$ and $|\Delta\alpha_X|\sim1.43$, also excludes the possibilities of the spectral regime transition, end of energy injection, or medium density drop to making such a break. The transition of the cooling frequency across the band predicts the change of the slopes is $\Delta\alpha = 0.25$ (Sari et al. 1998). The cessation of the energy injection process observed in long GRBs typically leads to $\Delta\alpha\sim 0.7$ (Nousek et al. 2006; Zhang et al. 2006; Liang et al. 2007). A steep drop in the density of the external medium is predicted to cause maximum changes of $\Delta\alpha \sim 0.4$ for density contrasts of $\sim10 $ (Nakar \& Granot 2007).

\subsection{Optical to X-ray afterglow spectrum}
\begin{figure}
\centering
\includegraphics[angle=0,scale=0.4]{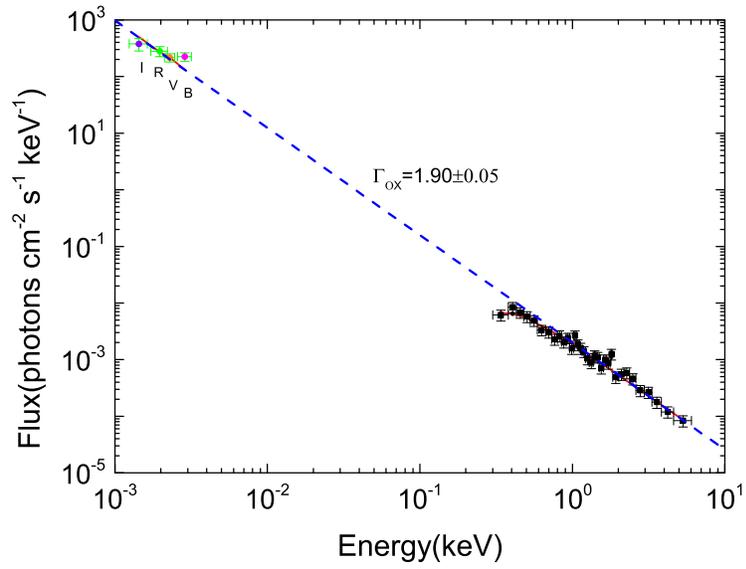}
\caption{The joint averaged optical-X-ray spectral energy distribution of the afterglow of GRB 140629A
derived from the time interval from 3084 seconds to 7000 seconds post the BAT trigger. The frequencies are expressed in the observer frame. The red solid line is our fitting curve of the broadband spectrum. The blue dashed line is
the intrinsic radiation spectrum derived from our fit.}
\label{obs_spec}
\end{figure}

Multi-wavelength data are available in the time interval from $T_0+3084$ seconds to $T_0+7000$ seconds. We construct the time-averaged broadband afterglow spectrum of GRB 140629A from this time interval.
Our optical data are corrected from the Galactic foreground extinction
with $A_I=0.012$, $A_R=0.018$, $A_V=0.022$ and $A_B=0.029$. The XRT spectrum is obtained from the XRT light curve and spectral repository (Evans et al. 2007, 2009). It is regrouped to ensure at least 20 counts
per bin using the tool ``grppha" in Xspec package. We fit the spectrum with a model $zdust*zpha*pha*powerlaw$ by using the Xspec package, where ``zdust" is for the dust extinction of the GRB 140629A host galaxy, ``zpha" and ``pha" are for the neutral hydrogen absorption of the GRB host galaxy and our Galaxy, respectively, and ``powerlaw" is a single power law function. We find that the optical extinction is negligible even when the extinction laws of the Galaxy, SMC and LMC were used in our fit. In addition, the neutral hydrogen absorption of the GRB host galaxy is also negligible. The absorption of the Galaxy with $N_H=9.32\times10^{19}$ cm$^{-2}$ is adequate to address the observed soft X-ray absorption. The spectrum is well fitted by our model with a  $\chi^2/dof=31.68 / 33$, where 33 is the degrees of freedom, as shown in Figure \ref{obs_spec}. The derived photon index is $\Gamma_{\rm OX}=-1.90\pm0.05$.

\subsection{Afterglow Light curve Fits with the External Shock Model}
In the framework of the standard afterglow model(e.g., Sari, Piran \& Narayan 1998; Huang et al. 2000; Yost et al. 2003), the multi-wavelength emission is radiated via the synchrotron process by relativistic electrons accelerated in forward shocks when the fireball propagates into the circumburst medium. For a constant density medium, the typical synchrotron emission
frequency, the cooling frequency and the peak spectral flux evolved with time are given by (Sari et al. 1998; Yost et
al. 2003; Fan \& Piran 2006; Zhang et al. 2007)
\begin{eqnarray}
\nu_m & = & 3.3 \times 10^{12} {\rm\ Hz}
\left(\frac{p-2}{p-1}\right)^2(1+z)^{1/2}\epsilon_{B,-2}^{1/2}
\epsilon_{e,-1}^{2}E_{K,52}^{1/2} t_d^{-3/2} \label{num}\\
\nu_c & = & 6.3 \times 10^{15} {\rm\ Hz}
(1+z)^{-1/2} (1+Y)^{-2} \epsilon_{B,-2}^{-3/2}E_{K,52}^{-1/2} n^{-1}
t_d^{-1/2} \label{nuc}\\ F_{\nu,\max} & = & 1.6 {\rm\ mJy}
(1+z)D^{-2}_{28}\epsilon_{B,-2}^{1/2}E_{K,52}n^{1/2}~
\label{Fnumax}
\end{eqnarray}
where $t_d$ is the observer's time in unit of days, $Y$ is the inverse Compton parameter, $D$ is the luminosity distance,
$\epsilon_e$ is the fraction of the shock energy in radiating electrons, $\epsilon_B$ is the fraction of the shock energy in magnetic fields,
$n$ is the medium density, $E_{\rm K,iso}$ is the isotropic kinetic energy, $p$ is the power-law index of the electron distribution, and $z$ is the redshift.

The standard GRB afterglow model discussed above is adopted in our modeling to derive the properties of the ejecta. We also consider the jet opening angle effect.  For GRB 140629A,  the optical flux decay slope  after the early peak is $\alpha_{\rm O,2}=-1.12\pm0.02$. During this decay segment, the derived photon index is $\Gamma_{\rm OX}=-1.90\pm0.05$. These values are consistent with the closure relation $\alpha=3\beta/2$ (Zhang et al., 2006), where $\beta=\Gamma+1$,   thus, we can infer that both the optical and X-ray afterglows are likely in the  synchrotron radiation spectral regime of $\nu_m < \nu_O < \nu_X < \nu_c$ in the slow cooling case for the ISM scenario (e.g., Sari et al. 1998; Zhang et al. 2006; Gao et al., 2013).
In this spectral regime the observed energy flux is given by
\begin{equation}
\nu F_{\nu}=F_{\nu,\max} (\nu_m/\nu_X)^{(p-1)/2}
\label{FnuX-2}
\end{equation}
and one can also infer $p=-2\beta_{\rm OX}+1\sim 2.80$, where $\beta_{\rm OX}=\Gamma_{\rm OX}+1$ (e.g., Sari et al. 1998; Zhang et al. 2006).

\begin{figure}
\centering
\includegraphics[angle=0,scale=0.6]{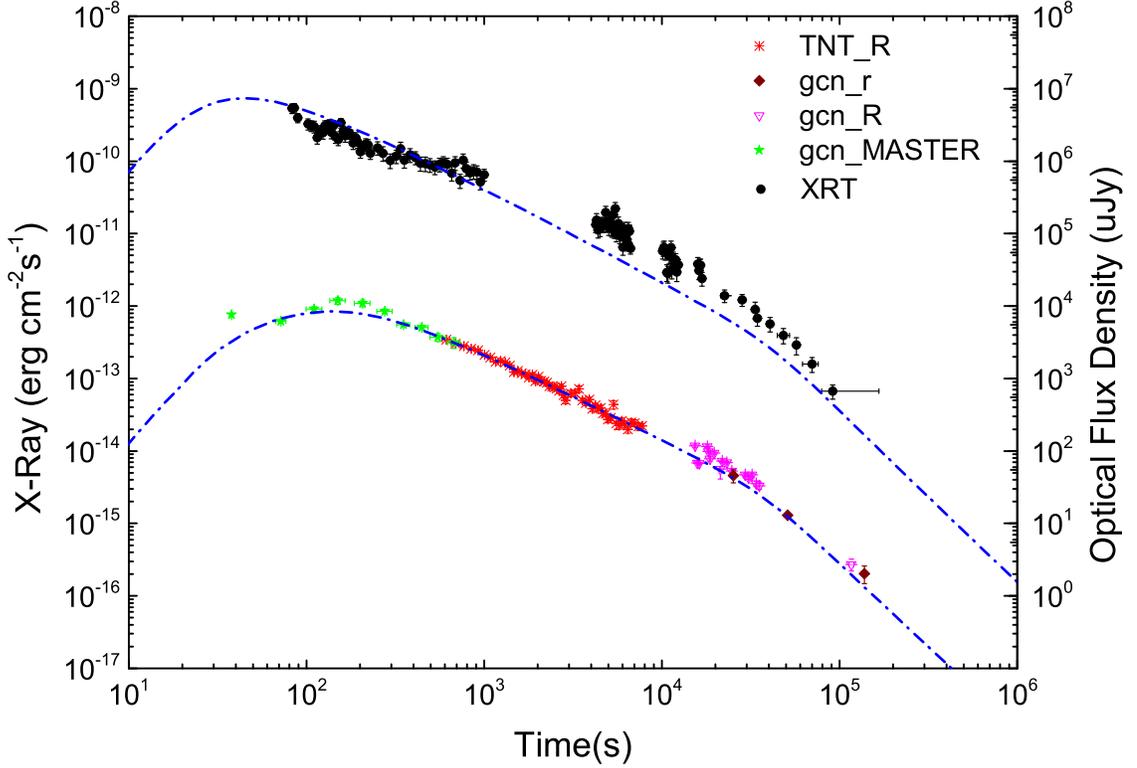}
\caption{The modeling of the optical and X-ray afterglow light curves with the synchrotron external shock model (blue dashed lines). The derived best micro physical parameters are : $\Gamma_0=315^{+44}_{-34}$, $\epsilon_e=(1.2\pm0.1)\times10^{-2}$, $\epsilon_B=(1.0\pm0.1)\times10^{-6}$, $n=60\pm9$  cm$^{-3}$, $E_{\rm K,iso}=(1.8\pm0.1)\times10^{55}$ erg, $p=2.72\pm0.07$ and $\theta_j=0.04^{+0.02}_{-0.01}$ rad.}
\label{model}
\end{figure}

\begin{figure}
\centering
\includegraphics[angle=0,scale=0.2]{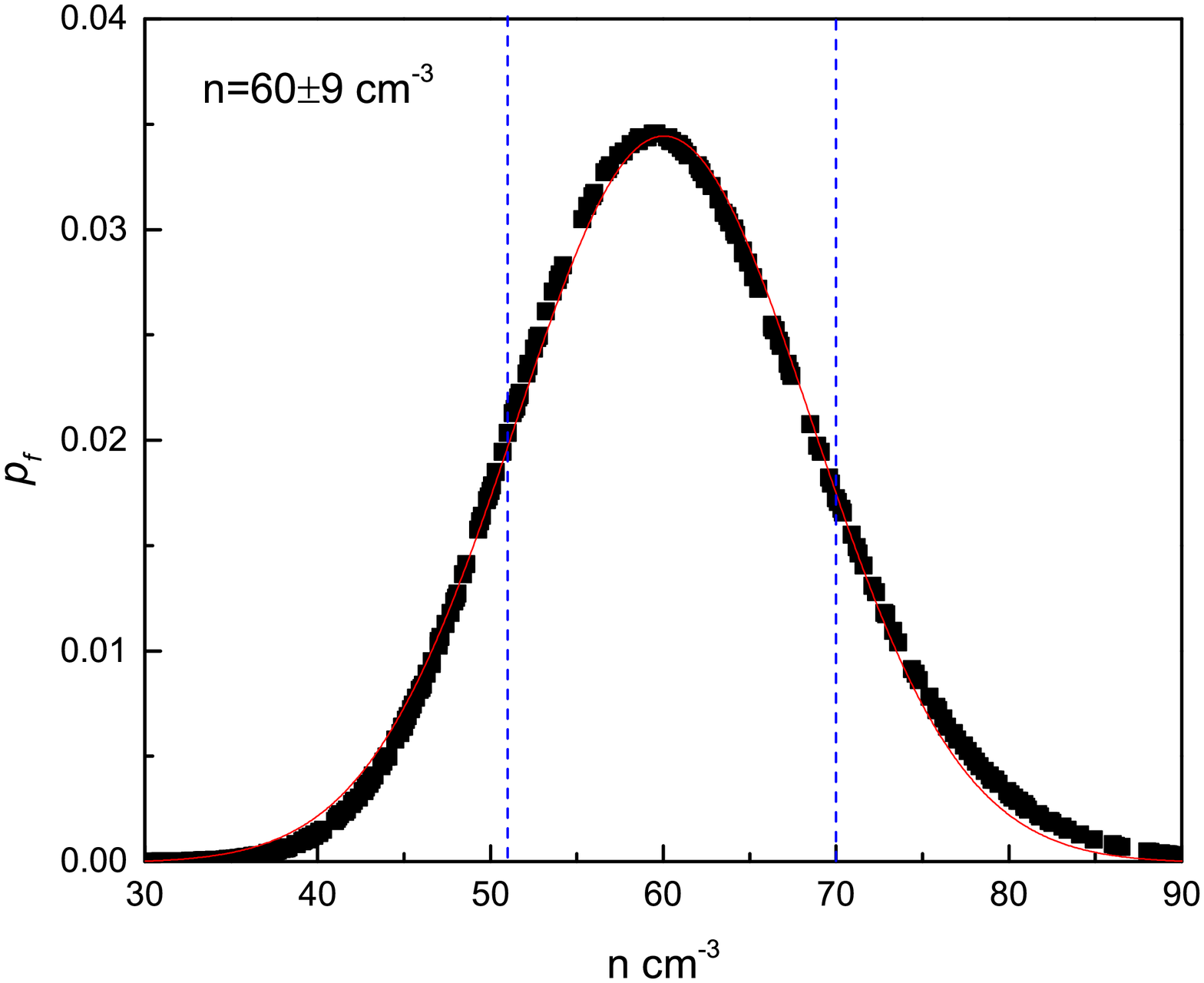}
\includegraphics[angle=0,scale=0.2]{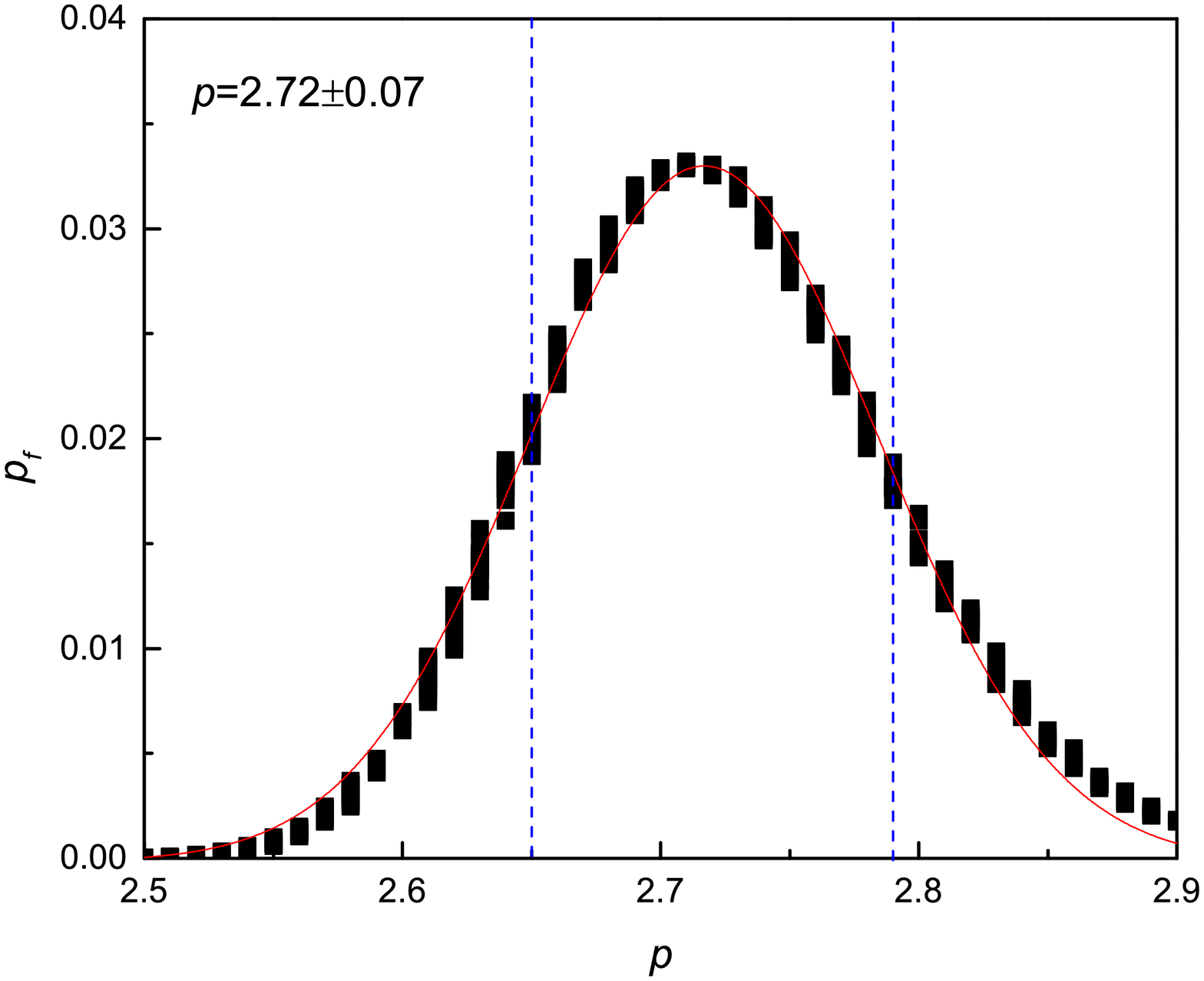}
\includegraphics[angle=0,scale=0.2]{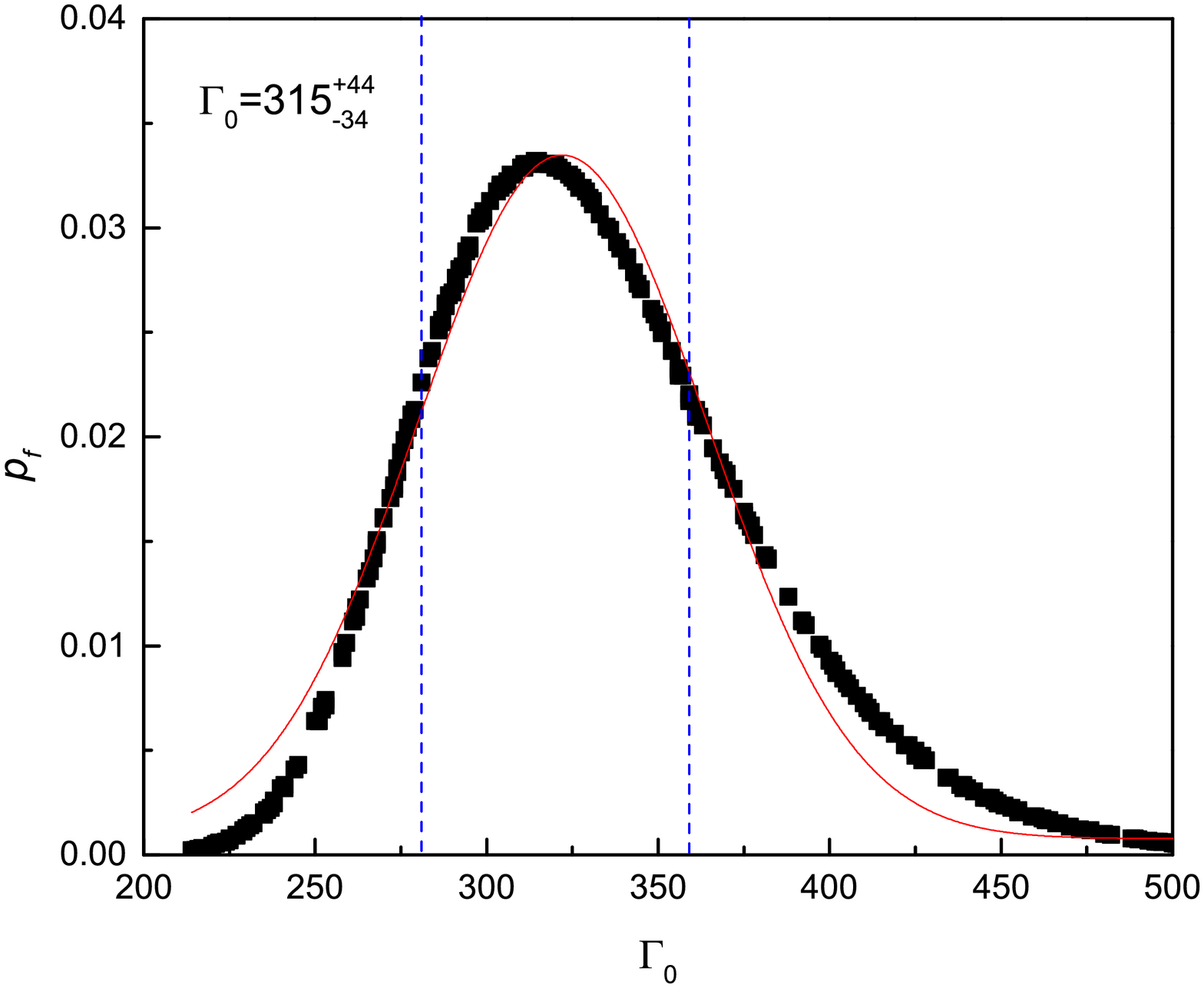}
\includegraphics[angle=0,scale=0.2]{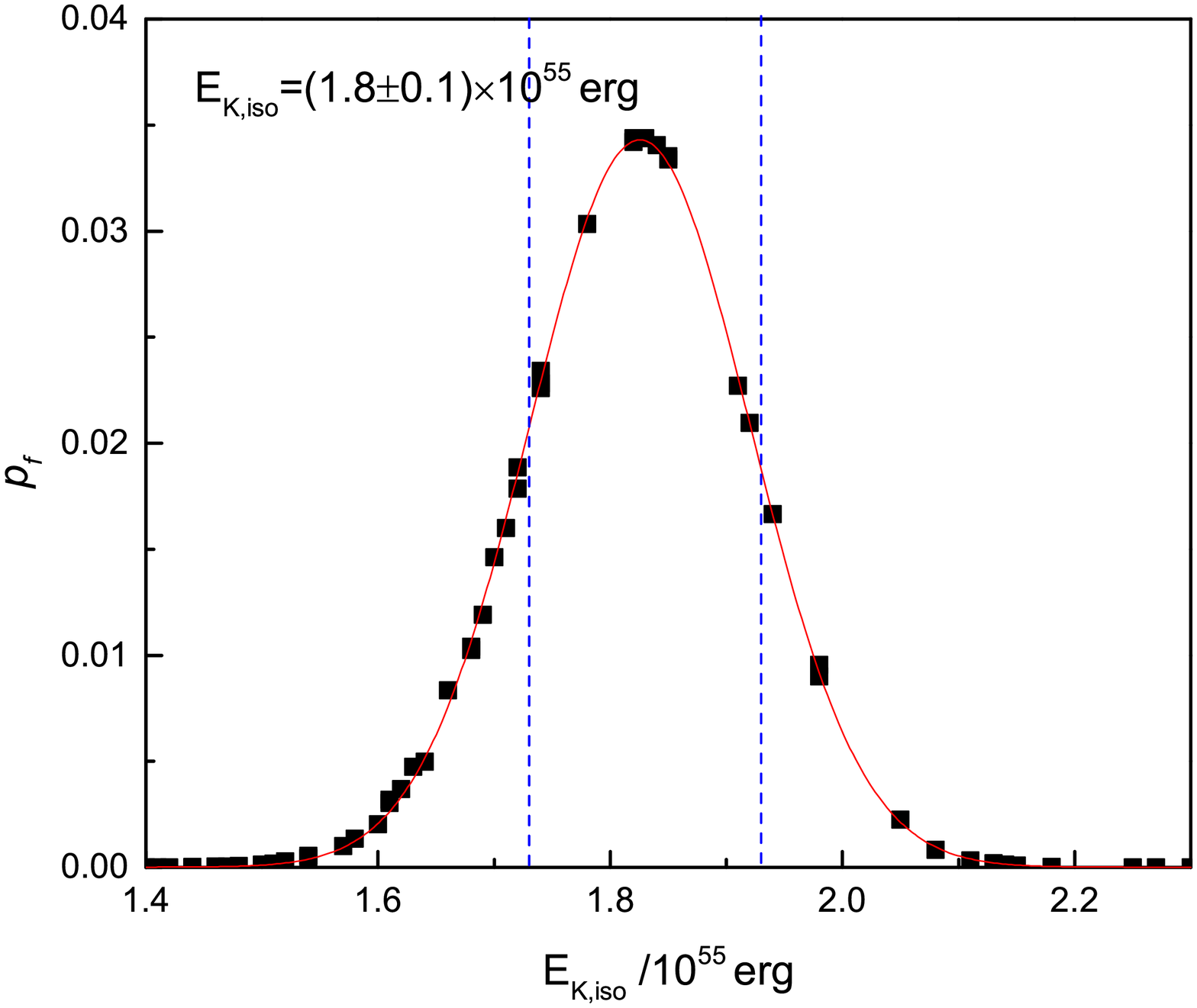}
\includegraphics[angle=0,scale=0.2]{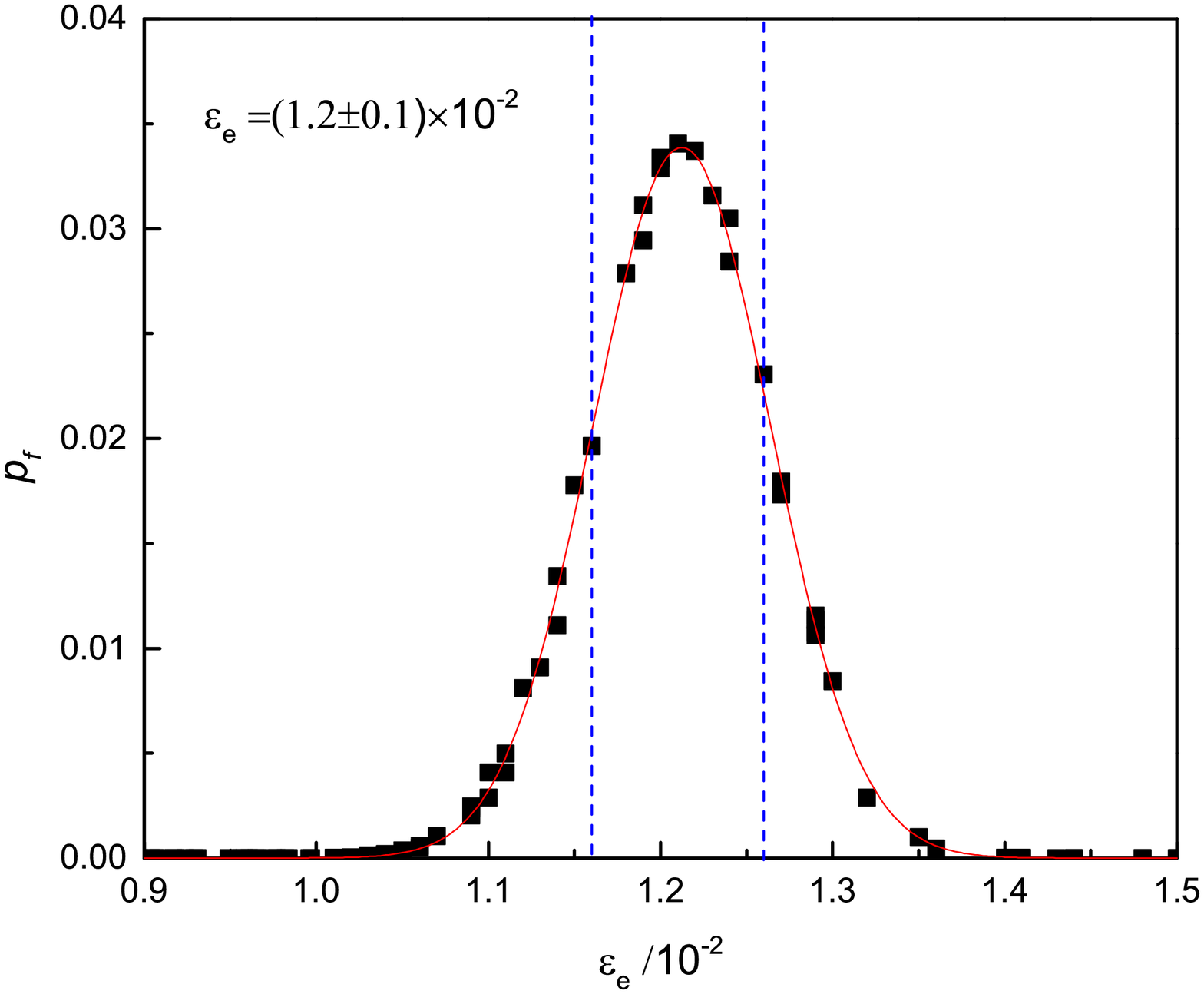}
\includegraphics[angle=0,scale=0.2]{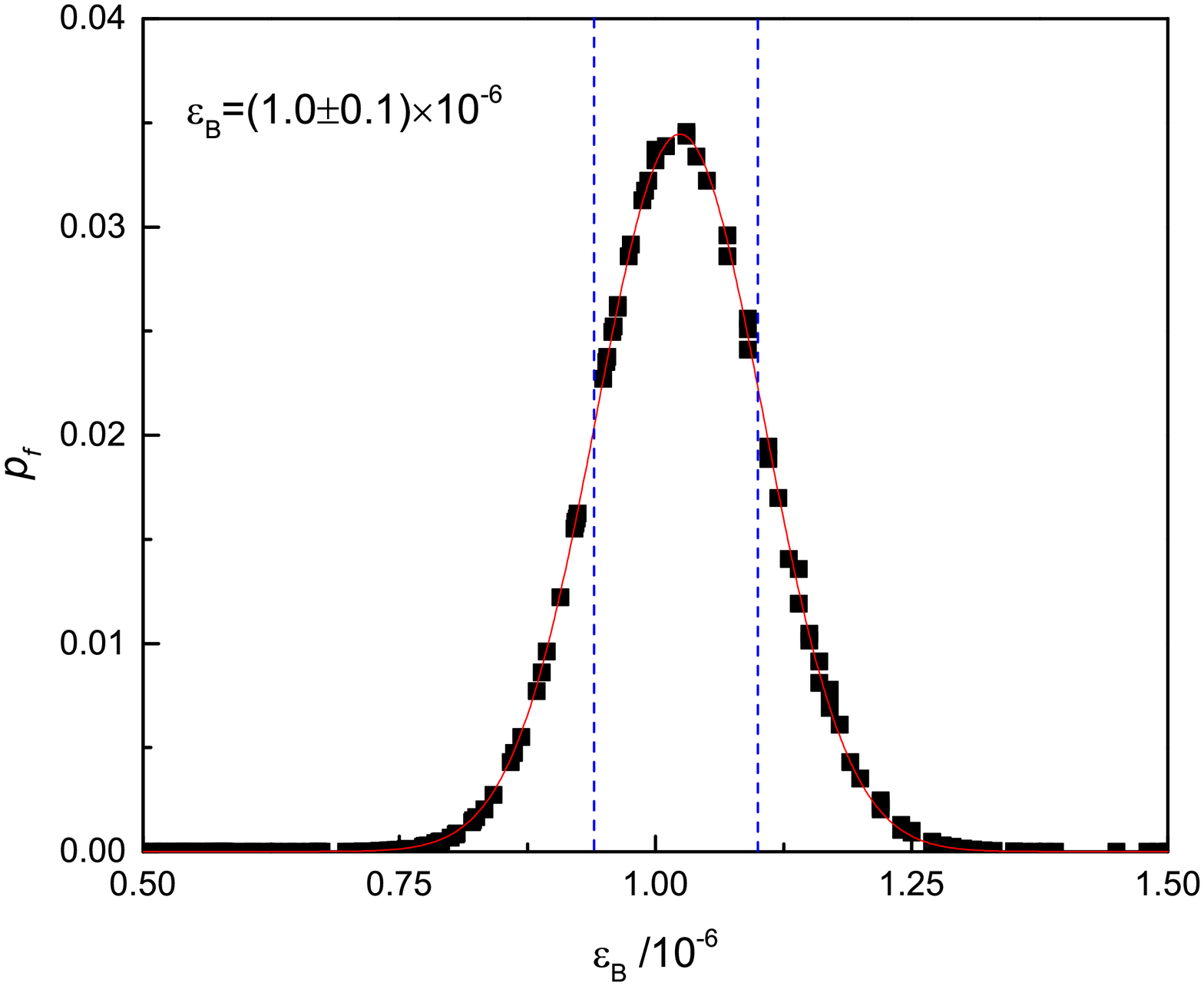}
\includegraphics[angle=0,scale=0.2]{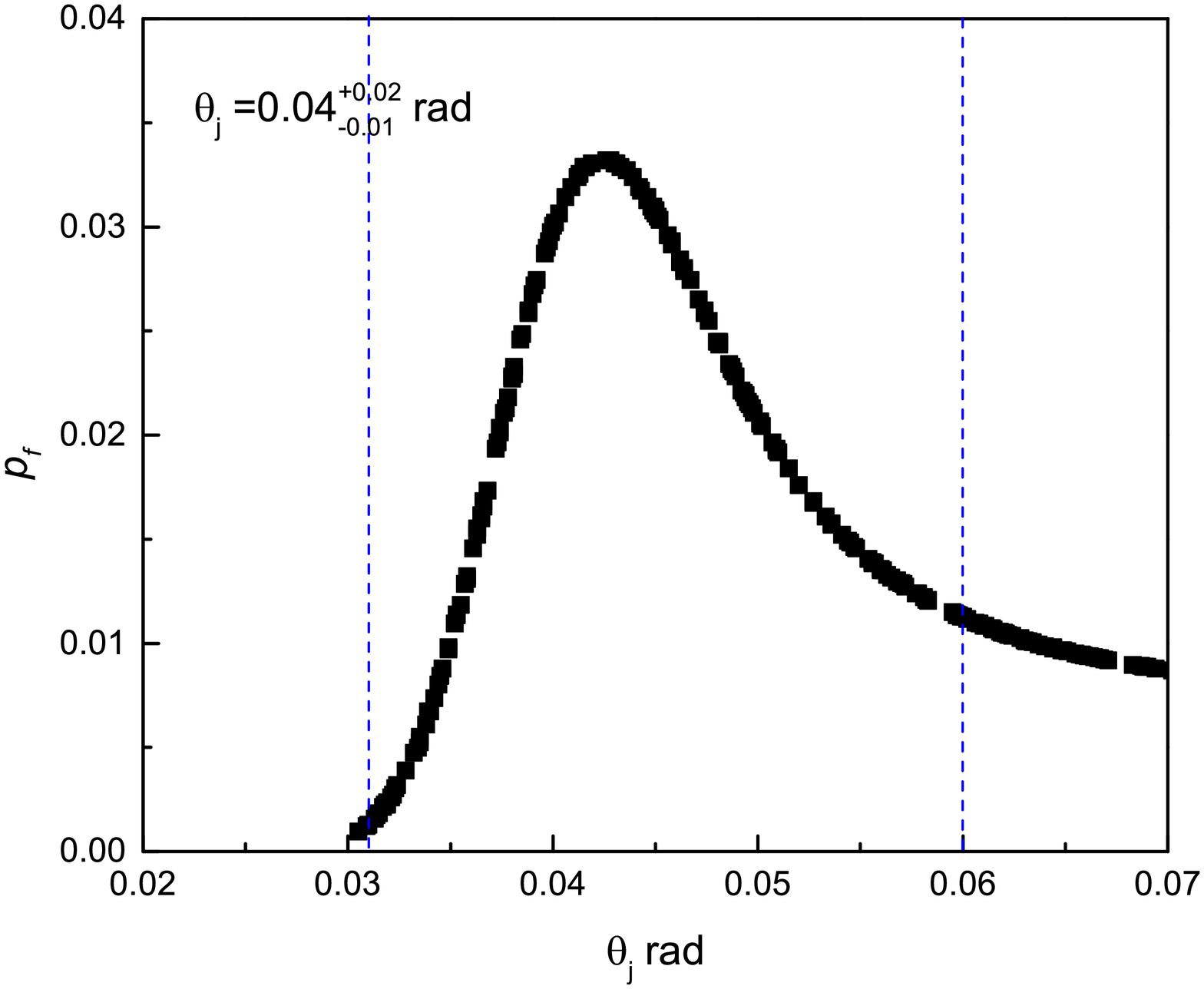}
\caption{The probability distributions of the afterglow model parameters along with our Gaussian function fits (solid red lines) for GRB 140629A. The dashed vertical lines mark the 1$\sigma$ confidence level of the parameters in this parameter set.  }
\label{Fitting_parameters}
\end{figure}

There are seven free parameters in our model, i.e., $\epsilon_e$, $\epsilon_B$, $n$, $E_{\rm K,iso}$, $p$, the jet opening angle ($\theta_j$), and the initial Lorentz factor ($\Gamma_0$). $Y$ is not an independent parameter, and whose treatment is the same as that in Fan \& Piran (2006).
To constrain these parameters, a Monte Carlo method is utillized to search for the best fit parameter set.
Following the technique and the procedure (Xin et al., 2016b),
the fitting results for GRB 140629A are illustrated in Figure \ref{model}, and the derived parameters and their probability distribution are shown in Figure \ref{Fitting_parameters}.  The $1\sigma$ confidence level estimated for the micro physical parameters are $\Gamma_0=315^{+44}_{-34}$, $\epsilon_e=(1.2\pm0.1)\times10^{-2}$, $\epsilon_B=(1.0\pm0.1)\times10^{-6}$, $n=60\pm9$  cm$^{-3}$, $E_{\rm K,iso}=(1.8\pm0.1)\times10^{55}$ erg, $\theta_j=0.04^{+0.02}_{-0.01}$ rad and $p=2.72\pm0.07$.

The derived $\epsilon_{\rm B}$ value is smaller than the typical values of $10^{-2}\sim 10^{-4}$ reported in the literature prior to the {\em Swift} mission era (e.g., Wijers \& Galama 1999; Panaitescu \& Kumar
2002; Yost et al. 2003; Panaitescu 2005).
Some recent statistical analysis working with both optical and X-ray afterglow data suggest a low  $\epsilon_{\rm B}$ value, i.e., $\sim 10^{-8} -10^{-3}$ (Santana et al., 2014; Japelj et al. 2014; Wang et al. 2015; Gao et al. 2015). Noting that both optical and X-ray afterglows are in the spectral regime $\nu<\nu_c$ in our modeling fit for GRB 140629A.
From equation (3), $\nu_c$ is proportional to $\epsilon_{B,-2}^{-3/2}n^{-1}t_d^{-1/2}$.  As time increases, $\nu_c$ is getting smaller.  One could also find that $\nu_c$ is more sensitive to $\epsilon_B$.
For GRB 140629A, the derived $n$ value is $60\pm9$  cm$^{-3}$.  In such dense medium,
the extremely low $\epsilon_B$ could ensure that both the optical and X-ray emission is still in the regime $\nu<\nu_c$ at late epoch.

 The model gives only a rough fit to the X-ray light curve. Note that our best empirical fit to the X-ray lightcurve derived a shallow decay segment with a slope of $-0.67\pm 0.02$ before $t<2\times 10^{3}$ seconds. However, we do not find similar feature in the optical lightcurve. We suspect that the shallow decaying behavior may partially resulted from the tail emission of the prompt gamma-rays of the second episode since the early X-ray emission was observed with XRT started at 93 seconds post the BAT trigger, being roughly at the end of this episode. 
Therefore, we do not consider any late energy injection in our modeling fit. In addition, significant flickering is also observed in the X-ray light curve. They are some residuals of late internal emission, which is difficult to depict their temporal details. 


 As shown in  Figure \ref{model}, the optical data at $t>300$ seconds are well represented by our model, but the optical data around the onset peak slightly deviates from it.
 Our empirical fit for GRB 140629A yields $\alpha_{\rm O,1}=0.92\pm 0.24$, which is much shallower than the predicted value of 3 by the model (e.g. Gao et al., 2013) in the thin shell case for a constant medium density. One possibility to explain the shallower rising slope is the temporal evolution of the medium density profile. Liang et al. (2013) found that the rising slope of the early afterglow onset is shallower than the prediction of a constant medium density for a large fraction of GRBs in their sample. They considered a circumburst medium density profile as
\begin{eqnarray}
\label{n} n = \left\{ \begin{array}{ll}
n_{0} \left(\frac{R}{R_{t}}\right)^{-k}, &  R <=R_t, \\
n_{0}, &  R > R_t,
\end{array} \right.
\end{eqnarray}
where $R_{\rm t}$ is the transition radius at which the medium turns into a constant density medium $n_0$. If the condition $R_{\rm t} \geq R_{\rm dec}$ is satisfied, where $R_{\rm dec}$ is the deceleration radius, the thin-shell external shock model gives a rising slope of $\alpha=3-k(p+5)/4$. They derived a typical $k$ value as $1$. As mentioned above, $p\sim2.7$, we then have $\alpha=1.07$ for GRB 140629A in this scenario. This value is consistent with that derived from our empirical fit within the error bar.
On the other hand, for the light curves post the peak time, it is also noticed that the decay index and the spectral slope  are consistent with the closure relation, $\alpha=3\beta/2$ (Zhang et al., 2006) in the  slow cooling case for the ISM scenario, indicating that the density profile of the medium after the peak is constant, $k=0$.
As a result,  the $k$ parameter before and after the peak time is changed from $\sim1$ to 0 . Consequently, the transition radius $R_t$ may be similar to the deceleration radius $R_{dec}$. 
This is also similar to that in GRB 121011A (Xin et al., 2016a).

Another possibility to interpret the shallow rising slope of the afterglow onset would be the contamination of the prompt optical emission or reverse shock emission. As mentioned in \S 4.1, the first optical data may be dominated by the prompt emission or the reverse shock emission (see also in  GRB 140512A; Huang et al., 2016). The early forward shock emission may contaminated by the prompt optical and/or reverse shock emission. 
If the  emission from the reverse shock and the forward shock at the early rising phase is comparable, the contamination effect would make a significant surplus in comparison with the prediction of the forward shock model

\section{Discussion}
\begin{figure}
\centering
\includegraphics[angle=0,scale=0.5]{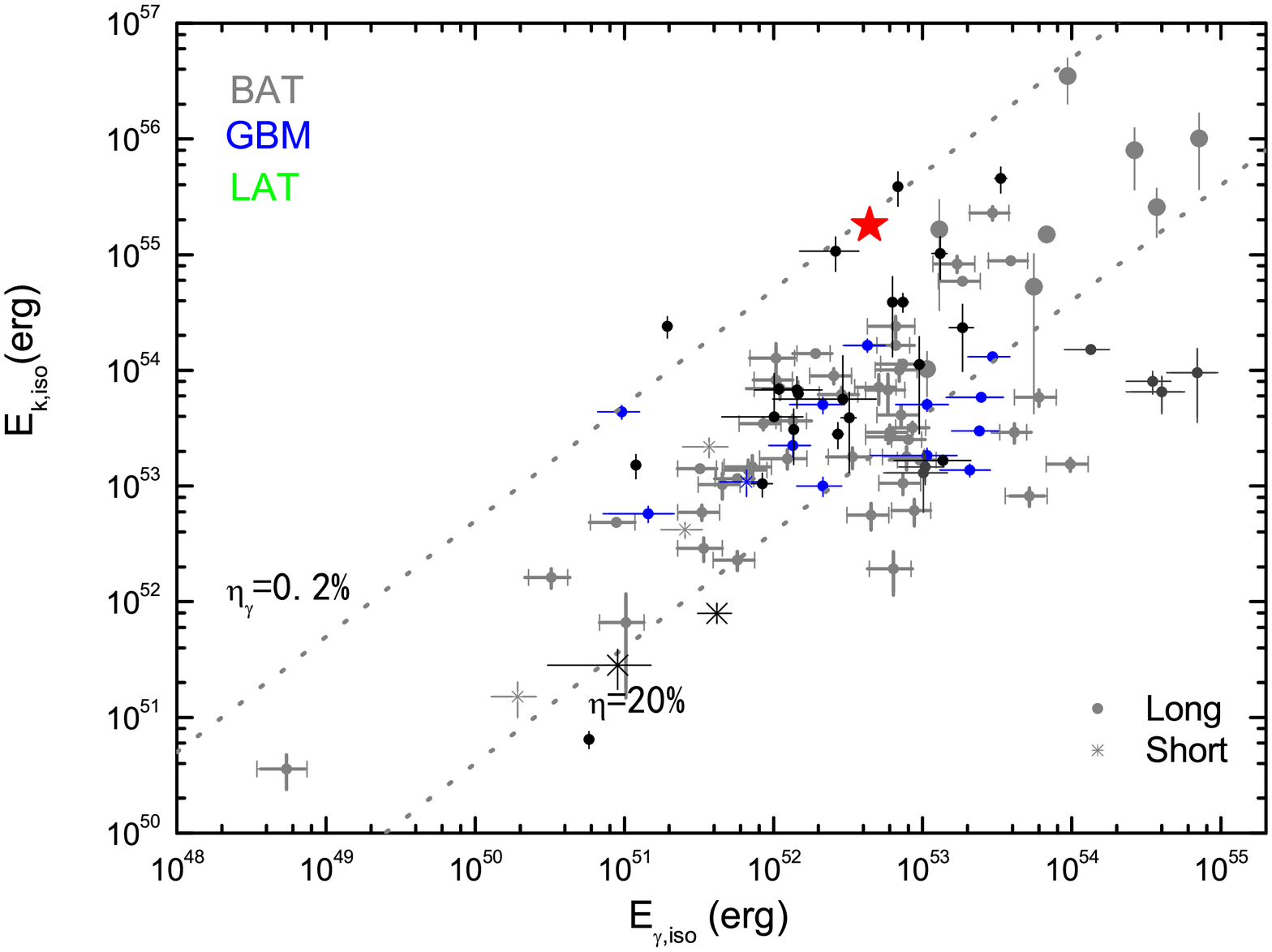}
\includegraphics[angle=0,scale=0.5]{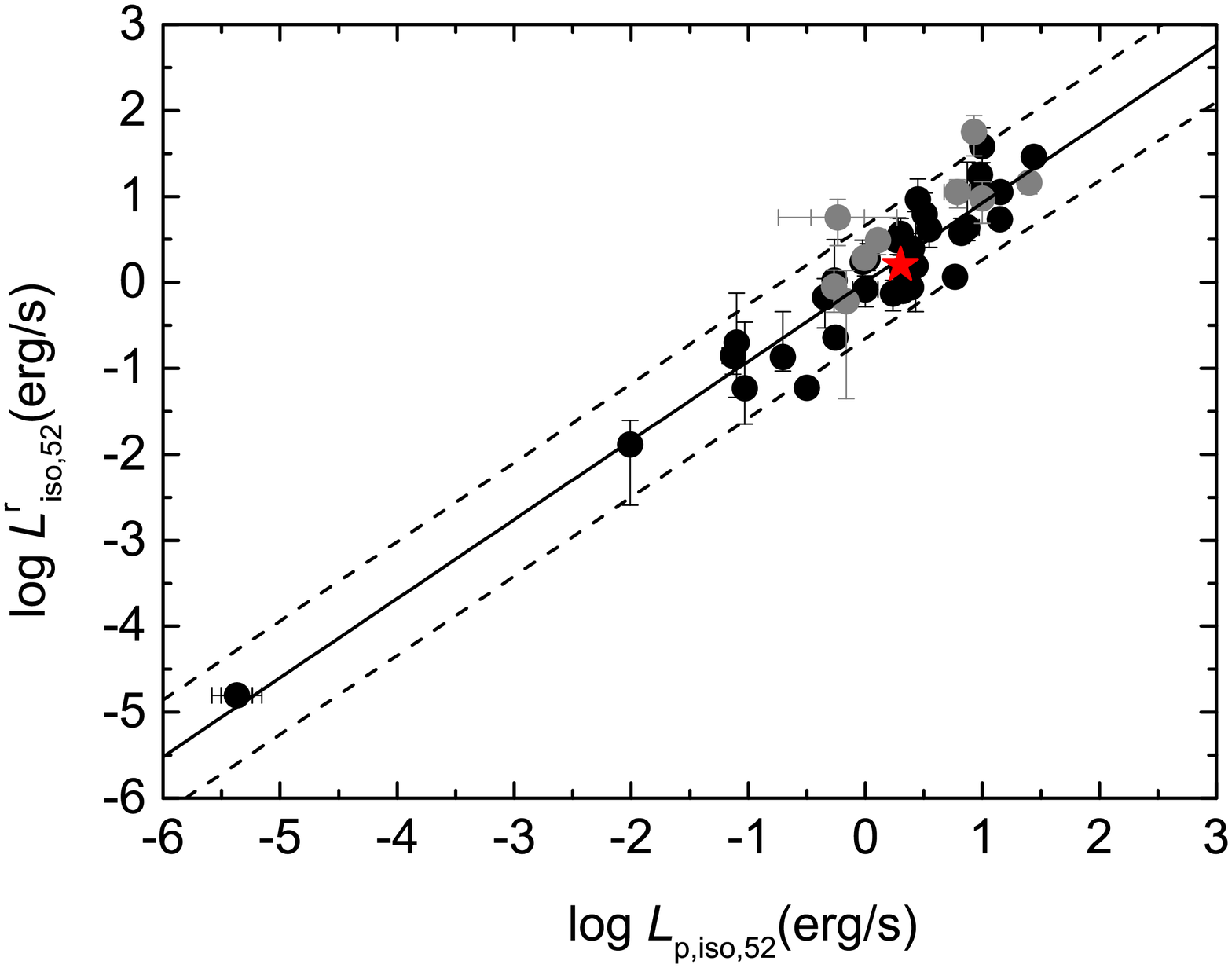}
\caption{{\em Upper---}The comparison of GRB 140629A (the red star) with typical GRBs in the $E_{\rm K,iso}-E_{\rm \gamma,iso}$ plane.  {\em Bottom}---The illustration of the consistency of GRB 140629A with the tight $L_{\rm \gamma, iso}-E_p^{'}-\Gamma_0$ relation (Liang et al. 2015).}
\label{relation}
\end{figure}

The GRB radiative efficiency ($\eta_\gamma$) is of theoretical interest since it may give some hints to the composition of the ejecta. With the measured redshift, $z=2.275$, the isotropic energy release $E_{\rm \gamma, iso}$ is estimated to be $4.4\times10^{52}$ erg using observed $S_\gamma$ in $20-10^4$ keV band. Therefore, we have $\eta_\gamma=E_{\rm \gamma, iso}/(E_{\rm K, iso}+E_{\rm \gamma, iso})=0.24\%$. It is extremely low in comparison with typical GRBs shown in Figure \ref{relation} (see also Zhang et al. 2007).
It was suggested that the GRB radiation efficiency is low in the keV-MeV band, if the radiation is produced by the internal shocks in collisions of ultra-relativistic matter shells (e.g., Kobayashi et al. 1997; Daigne \& Mochkovitch 1998; Kumar 1999; Panaitescu et al. 1999)\footnote{The radiation efficiency may be much higher ($\sim 40\%$) when the inner engine produces fireball shells with comparable energies but with very different Lorentz factors (Kobayashi et al. 1997).}. The derived low efficiency is consistent with the prediction of the standard internal shock model.

For GRB 140629A, our analysis suggests that the optical and X-ray afterglows are from a narrow jet ($\theta_j=0.04^{+0.02}_{-0.01}$ rad) with a low-$\epsilon_B$ [$(1.0\pm 0.01)\times 10^{-6}$] in a dense medium ($n=60$ cm$^{-3}$). In addition, the radiation efficiency of GRB 140629A is extremely low. We test whether or not it satisfies various empirical relations reported in the literature derived from observations of the prompt gamma-ray phase and the multi-wavelength afterglows.
By estimating the jet opening angle with a jet-like break time $t_j$ in late multi-wavelength light curves, Ghirlanda et al. (2004a) derived a tight correlation between geometrically-corrected jet energy $E_{\gamma,j}$  and the peak energy $E^{'}_p$ of $\nu f_{\nu}$ spectrum in the burst frame, i.e., $E_p^{'}=267.0(E_{\gamma,j}/4.3\times 10^{50} {\rm ergs})^{0.706\pm0.047}$. The $E_p^{'}$ value inferred from the Ghirlanda relation is 46 keV for GRB 140629A, which is definitely inconsistent with the data, i.e., $E_p^{'} \sim E_p \times (1+z) \sim 283$ keV.
Liang \& Zhang (2005) derived an empirical relation among $E_{\gamma, {\rm iso}}$, $E_p^{'}$, and the jet break time  ($t^{'}_j$) in the burst frame, i.e., $E_{\gamma, {\rm{iso}}}/10^{52} {\rm
ergs}=(0.85\pm0.21)\times (E^{'}_{\rm {p}}/{\rm 100 \
keV})^{1.94\pm0.17}\times(t^{'}_{\rm{b}}/1{\rm day})^{-1.24\pm0.23}$. Based on the relation, the isotropic energy $E_{\gamma, {\rm iso}}=7.9\times 10^{53}$ ergs is obtained, which is larger than the observed one with more than one order of magnitude. These results suggest that GRB 140629A does not follow the two relations (Ghirlanda et al., 2004a; Liang \& Zhang 2005), although both tight  correlations have been used for measuring the cosmological parameters with GRBs (e.g., Dai et al. 2004; Ghirlanda et al. 2004b; Liang \& Zhang 2005;  Wang, Dai \& Liang 2015).
Note that the observed jet break time of GRB 140629A is much earlier, hence the inferred $\theta_j$ is much lower than those of the GRBs used to derive these relations (e.g., Frial et al. 2001; Bloom et al. 2003). It is unclear whether the violation of GRB 140629A is due to the selection effect or other physical reasons. For example, two-component jet models composed of a narrow and a wide components have been proposed to explain the data of some GRBs (e.g., Huang et al. 2004; Racusin et al. 2008). In these cases, the high-energy emission was proposed to be emitted by the narrow jet. However, one cannot exclude the possibility that the observed gamma-ray energy would be dominated by the wide jet component under certain conditions. Meanwhile, the early break time for GRB 140629A is likely to be due to the jet effect of the narrow jet component but not the wide one. If it is the case, the inconsistency between the jet energy and the opening angle would result in this violation of  GRB 140629A.
Liang et al. (2015) discovered a tight empirical correlation among $L_{\rm iso}$, $E^{'}_{\rm p}$ and $\Gamma_0$ to reveal the direct connection between the gamma-ray and afterglows, \begin{equation}L_{\rm iso, 52}=10^{-6.38\pm 0.35}{(E^{'}_{\rm p}/{\rm keV})}^{1.34\pm 0.14}\Gamma_{0}^{1.32\pm 0.19.}\end{equation}  Based on the equation above, we get $L_{\rm iso, 52}=1.60^{+0.32}_{-0.30}$ for GRB 140629A, where the error is calculated with the uncertainties of $E^{'}_p$ and $\Gamma_0$ only. The derived $L_{\rm iso, 52}$ is well consistent with the observed one, $2.0\times10^{52}$ erg/s, as shown in Figure \ref{relation}. Note that the initial Lorentz factor of the ejecta $\Gamma_0$ is sensitive to the deceleration time (the peak time of the onset bump), but not strongly related to the jet break time. The onset of the afterglow  bump is usually bright (Liang et al. 2010, 2013; Li et al. 2012; Wang et al. 2013), and it is easier to be identified than the jet break time from an observed light curve\footnote{The jet-break is usually detected in late optical afterglow light curve. It is dim and also contaminated by emission from the host galaxy and/or associated supernovae (e.g., Li et al. 2012). It is also an issue for identifying an observed jet break as the narrow or the wide component in the case of two-component jet.}. The consistency of GRB 140629A with the $L_{\rm iso}-E_p^{'}-\Gamma_0$ may suggest that this relation would be a more robust one than the Ghirlanda relation and Liang-Zhang relation since it is not sensitive to the jet opening angle $\theta_j$.

\section{Conclusions}
We have presented our optical photometric and spectroscopic observations of GRB 140629A with the TNT telescope and the 2.16 m telescope at Xinglong Observatory. The redshift of GRB 140629A of $z=2.275\pm0.043$ is measured through the metal absorption lines from our spectroscopic data. With the equivalent widths of the lines CIV$\lambda1549$ and CII$\lambda$1335 measured from our Gaussian fits to the line profiles, we obtain the ratio of their equivalent widths as $2.79\pm0.49$, indicating a high ionization level of the surrounding environment due to the GRB's radiation at the early phase after the burst. The optical-to-X-ray afterglow spectrum is jointly fitted with a single power-law function, yielding a photon index of $-1.90\pm 0.05$. The optical extinction and the neutral hydrogen absorption of the GRB host galaxy are negligible. We fit the optical and X-ray afterglow light curves with the forward shock model and find that the model can well represent the observed light curves with the following parameter set, i.e., $\Gamma_0=315^{+44}_{-34}$, $\epsilon_e=(1.2\pm0.1)\times10^{-2}$, $\epsilon_B=(1.0\pm0.1)\times10^{-6}$, $n=60\pm9$  cm$^{-3}$, $E_{\rm K,iso}=(1.8\pm0.1)\times10^{55}$ erg, $p=2.72\pm 0.07$  and $\theta_j=0.04^{+0.02}_{-0.01}$ rad. The  extremely low GRB radiation efficiency derived from our analysis agrees well with the prediction of the baryonic-dominated jet models. The extremely small opening angle makes GRB 140629A confidently violate the Ghirlanda relation and Liang-Zhang relation. However, it still agrees well with the $L_{iso}-E_p^{'}-\Gamma_0$ relation.

\section{Acknowledgement}
We very appreciate the valuable comments from the anonymous referee and Turpin Damien.
We also thank Bing Bing Zhang for his discussion on the X-ray data analysis.
This work is supported by the National Basic Research Program of China (973 Program,
grant No. 2014CB845800), the National Natural Science Foundation of China (Grant No.
11533003, U1731239, and U1331101). EWL is also support by a special funding from the Guangxi Science Foundation for Guangxi distinguished professors (Bagui Yingcai \& Bagui Xuezhe; 2017AD22006). JW is supported by the National
Natural Science Foundation of China under grants 11473036 and 11273027.
We acknowledge the support of the staff of the Xinglong 2.16m telescope. This work was partially supported by the Open Project Program of the Key Laboratory of Optical Astronomy, National Astronomical Observatories, Chinese Academy of Sciences.
This work made use of data supplied by the UK Swift Science Data Centre at the University of Leicester.

\clearpage

\begin{deluxetable}{cccccccccc}
\tabletypesize{\footnotesize}
\tablewidth{0.99\textwidth}
\center
\tablecaption{Optical Afterglow Photometry Log$^*$ of GRB 140629A by TNT}
\tablehead{
\colhead{$T-T_0$ } &
\colhead{Exposure} &
\colhead{Filter} &
\colhead{Mag} &
\colhead{$\sigma$}  &
\colhead{$T-T_0$} &
\colhead{Exposure}   &
\colhead{Filter} &
\colhead{Mag} &
\colhead{$\sigma$}
}
\startdata
3087  &  40   &  $B$  &  17.47  &  0.18  &        2722  &  60   &  $R$  &  16.46  &  0.08  \\
2950  &  40   &  $V$  &  17.05  &  0.15  &        2800  &  60   &  $R$  &  16.76  &  0.10  \\
3308  &  60   &  $V$  &  17.17  &  0.11  &        2878  &  60   &  $R$  &  16.94  &  0.11  \\
3471  &  60   &  $V$  &  17.40  &  0.12  &        3077  &  40   &  $R$  &  16.71  &  0.11  \\
3633  &  60   &  $V$  &  17.42  &  0.10  &        3251  &  40   &  $R$  &  16.69  &  0.10  \\
3795  &  60   &  $V$  &  17.33  &  0.08  &        3414  &  40   &  $R$  &  16.55  &  0.11  \\
3957  &  60   &  $V$  &  17.41  &  0.08  &        3576  &  40   &  $R$  &  16.94  &  0.08  \\
4120  &  60   &  $V$  &  17.55  &  0.09  &        3738  &  40   &  $R$  &  17.04  &  0.08  \\
4272  &  60   &  $V$  &  17.58  &  0.09  &        3901  &  40   &  $R$  &  16.91  &  0.07  \\
4424  &  60   &  $V$  &  17.65  &  0.09  &        4063  &  40   &  $R$  &  17.23  &  0.09  \\
4577  &  60   &  $V$  &  17.80  &  0.11  &        4215  &  40   &  $R$  &  17.12  &  0.07  \\
4729  &  60   &  $V$  &  17.84  &  0.12  &        4368  &  40   &  $R$  &  17.23  &  0.08  \\
4881  &  60   &  $V$  &  17.98  &  0.12  &        4520  &  40   &  $R$  &  17.21  &  0.08  \\
5044  &  80   &  $V$  &  17.81  &  0.10  &        4672  &  40   &  $R$  &  17.39  &  0.09  \\
5216  &  80   &  $V$  &  17.97  &  0.11  &        4824  &  40   &  $R$  &  17.40  &  0.09  \\
5408  &  80   &  $V$  &  17.88  &  0.09  &        4977  &  40   &  $R$  &  17.60  &  0.11  \\
5600  &  80   &  $V$  &  17.81  &  0.10  &        5149  &  40   &  $R$  &  17.55  &  0.10  \\
5813  &  80   &  $V$  &  18.13  &  0.13  &        5341  &  40   &  $R$  &  17.09  &  0.14  \\
6025  &  80   &  $V$  &  18.15  &  0.14  &        5534  &  40   &  $R$  &  17.73  &  0.19  \\
6247  &  100  &  $V$  &  18.04  &  0.12  &        5736  &  60   &  $R$  &  17.81  &  0.12  \\
6529  &  100  &  $V$  &  17.93  &  0.12  &        5948  &  60   &  $R$  &  17.65  &  0.10  \\
611   &  60   &  $R$  &  14.85  &  0.02  &        6160  &  60   &  $R$  &  17.76  &  0.11  \\
690   &  60   &  $R$  &  14.99  &  0.03  &        6433  &  80   &  $R$  &  17.95  &  0.12  \\
768   &  60   &  $R$  &  15.07  &  0.03  &        6715  &  80   &  $R$  &  17.70  &  0.12  \\
846   &  60   &  $R$  &  15.16  &  0.03  &        7027  &  100  &  $R$  &  17.73  &  0.12  \\
924   &  60   &  $R$  &  15.22  &  0.03  &        7407  &  120  &  $R$  &  17.84  &  0.13  \\
1003  &  60   &  $R$  &  15.37  &  0.03  &        7729  &  200  &  $R$  &  17.83  &  0.09  \\
1081  &  60   &  $R$  &  15.47  &  0.03  &        3367  &  40   &  $I$  &  16.56  &  0.10  \\
1159  &  60   &  $R$  &  15.61  &  0.04  &        3529  &  40   &  $I$  &  16.55  &  0.08  \\
1237  &  60   &  $R$  &  15.59  &  0.03  &        3691  &  40   &  $I$  &  16.63  &  0.09  \\
1315  &  60   &  $R$  &  15.64  &  0.04  &        3854  &  40   &  $I$  &  16.53  &  0.07  \\
1393  &  60   &  $R$  &  15.75  &  0.05  &        4016  &  40   &  $I$  &  16.75  &  0.08  \\
1471  &  60   &  $R$  &  15.97  &  0.05  &        4173  &  30   &  $I$  &  16.70  &  0.09  \\
1549  &  60   &  $R$  &  15.93  &  0.04  &        4326  &  30   &  $I$  &  16.81  &  0.09  \\
1628  &  60   &  $R$  &  16.01  &  0.04  &        4478  &  30   &  $I$  &  16.94  &  0.10  \\
1706  &  60   &  $R$  &  16.04  &  0.05  &        4630  &  30   &  $I$  &  16.83  &  0.10  \\
1784  &  60   &  $R$  &  16.15  &  0.05  &        4783  &  30   &  $I$  &  17.10  &  0.12  \\
1862  &  60   &  $R$  &  16.06  &  0.05  &        4935  &  30   &  $I$  &  16.93  &  0.11  \\
1940  &  60   &  $R$  &  16.29  &  0.06  &        5107  &  30   &  $I$  &  17.07  &  0.12  \\
2018  &  60   &  $R$  &  16.12  &  0.05  &        5482  &  50   &  $I$  &  17.29  &  0.11  \\
2096  &  60   &  $R$  &  16.27  &  0.06  &        5674  &  50   &  $I$  &  17.26  &  0.12  \\
2175  &  60   &  $R$  &  16.30  &  0.07  &        5886  &  50   &  $I$  &  17.22  &  0.11  \\
2253  &  60   &  $R$  &  16.34  &  0.07  &        6099  &  50   &  $I$  &  17.46  &  0.14  \\
2331  &  60   &  $R$  &  16.46  &  0.08  &        6346  &  80   &  $I$  &  17.35  &  0.10  \\
2409  &  60   &  $R$  &  16.47  &  0.07  &        6628  &  80   &  $I$  &  17.29  &  0.11  \\
2487  &  60   &  $R$  &  16.49  &  0.08  &        6930  &  80   &  $I$  &  17.21  &  0.12  \\
2566  &  60   &  $R$  &  16.53  &  0.07  &        7290  &  100  &  $I$  &  17.84  &  0.23  \\
2644  &  60   &  $R$  &  16.61  &  0.09  &              &       &       &         &        \\
\enddata
\tablenotetext{*}{The reference time $T_0$ is {\em Swift} BAT burst trigger time. "$T-T_0$" is the middle time in second. "Exposure" is the exposure time in second. "$\sigma$" means the uncertainty of magnitude.
All data are calibrated by nearby $SDSS$ reference stars.
All data are not corrected for the Galactic extinction (which is $E_{B-V}=0.01$, Schlegel et al.1998).}
\label{Tab1}
\end{deluxetable}

\clearpage
\begin{table}
\caption{Measured EWs in the rest-frame from the afterglow spectrum of GRB\,140629A}
\centering
\begin{tabular}{ccc}
\hline\hline
Line identification & Central wavelength & EWs  \\
& \AA & \AA\\
\hline
CII  & 1335 &  $1.80\pm0.31$ \\
SiIV & 1394 & $1.74\pm0.25$ \\
SiIV & 1403 & $2.01\pm0.25$ \\
CIV & 1549 & $5.02\pm0.21$ \\
AlII & 1671 & $1.28\pm0.30$ \\
\hline
\end{tabular}
\label{Tab2}
\end{table}

\begin{table}
\caption{The fitting results of the multi-wavelength afterglow light curves of GRB\,140629A. Note that the value of $\chi^2/dof$ for optical data labeled by a star (*) in this table is slight large, due to the bad fitting for the late optical data. If the fitting is only made to the optical data before $10^{4} $ sec after the burst trigger time, the value of $\chi^2/dof$ would be  $\sim$2.20.  }
\centering
\tabletypesize{\footnotesize}
\begin{tabular}{cccccccc}
\hline\hline
Band & $\alpha_1$ & $\alpha_{2}$ & $\alpha_{3}$ & $t_{p}(s) $ & $t_{b}(ks)$ & $t_{j}(ks) $  & $\chi^2/dof$\\
\hline
Optical  & 0.92$\pm$0.24 & -1.12$\pm$0.02 & -2.35$\pm$0.24  & 179$\pm$16 & - & 37.2(fixed) & 8.90$^{*}$ \\
X-ray    &  -0.67$\pm$0.02  & -1.31$\pm$0.08 & -2.76$\pm$0.40 & - & 2(fixed)  & 37.2$\pm9.1$  & 1.06 \\
\hline
\end{tabular}
\label{Tab3}
\end{table}



\end{document}